# Spiral Thermal Waves Generated by Self-Propelled Camphor Boats

Alla Vilk, Irina Legchenkova, Mark Frenkel and Edward Bormashenko[*]

*Ariel University, Engineering Faculty, Chemical Engineering Department,*

*P.O.B. 3, 407000, Ariel, Israel*

Corresponding author: E-Mail: edward@ariel. ac.il

**ABSTRACT**

Spiral thermal surface waves arising from self-propulsion of the camphor-driven objects are reported. Spiral thermal waves were registered for the dissolution and evaporation-guided self-propulsion. Soluto-capillarity is accompanied by thermo-capillarity under self-propulsion of camphor boats. The jump in the surface tension due to the soluto-capillarity is much larger than that due to the thermo-capillarity. The observed thermal effect is related to the adsorption of camphor molecules at the water/vapor interface. The observed spirals are shaped as Archimedean ones.

The fascinating phenomenon of spiral waves was registered within various experimental contexts, including formation of spiral galaxies[1], rotating liquids[2-4] and insect population[5]. The particular case of spiral waves are thermal waves, in which spirals are formed with the 2D thermal fields, inherent for example for the Belousov-Zhabotinsky reaction.[6-9] We demonstrate spiral thermal waves emerging from the self-propulsion of camphor boats, floating on water. The phenomenon of self-propulsion attracted much attention of investigators during the past decade.[10-15] Self-propulsion was successfully exploited recently for micro-robotics applications as well as for generation of electrical power.[16-18] One of the most studied self-propelled object is the so called "camphor boat" studied first by Lord Rayleigh.[19] The motion of camphor boats is usually related to the soluto-capillary interfacial (Marangoni) flows.[12,13, 20-23] Camphor decreases the surface tension of water, and the boat is driven by difference in the surface tension around it.[20,21] Obviously motion becomes possible due to breaking of symmetry of the surface tension field, surrounding the camphor boat.[20-25] Breaking of symmetry may arise from the non-uniform dissolution of camphor. We demonstrate that thermal Marangoni flows accompanying the self-propulsion of camphor boats may be not negligible and also contribute to the self-propulsion. Moreover, the self-propulsion of camphor boats gives rise to the spiral thermal fields, resembling those inherent for the Belousov-Zhabotinsky reaction.[6-9]

The experimental system is shown schematically in **Figure 1a-b**. Generally self-propulsion of the camphor-driven objects is possible under two scenarios, namely under evaporation and dissolution of camphor.[19-21, 25] Both of the regimes were studied experimentally. When the self-propulsion was guided by dissolution of camphor, the grains were placed directly on the water surfaces, as shown in **Figure 1a**. Special rotator depicted in **Figure 1b** was designed for the study of the evaporation-guided self-propulsion, as depicted in **Figure 1b**. Camphor was placed in the polymer tubing, with a diameter of 1 mm.

Vessels of various shapes were used for the study of the self-propulsion: Petri dishes (Corning Crystal Polystyrene) with the diameters of 140 mm and 85mm, square polystyrene vessels with dimensions of 210x210 mm and 110x180 mm. The de-ionized water, used as supporting liquid, was purified by a synergy UV water purification system from Millipore SAS (France) and its specific resistivity was $\hat{\rho} = 18.2 \text{M}\Omega \times \text{cm}$ at 25ºC. The vessels were filled with de-ionized water, the height of the supporting liquid $h_l$ was 7±1 mm. Camphor grains with dimensions of (4±1)x(4±1)x(6±2) mm were prepared from Camphor (96%), supplied by Sigma-Aldrich. Temperature of water varied within the range 25–35°C. All the experiments were performed under atmospheric pressure. The self-propelled motion of camphor grains was visualized with Therm-App TAS19AQ-1000-HZ thermal camera.

Thermal imaging of the water/vapor interface registered spiral thermal waves such as those depicted in **Figure 2a-c**. Spiral waves were observed in the circular and rectangular vessels. Clockwise and counter-clockwise rotation of water was observed within the spiral thermal waves generated by camphor grains and rotors driven by evaporation of camphor. Evaporation-induced rotation may be controlled by the shape of polymer tubing. It should be emphasized that the spiral thermal waves emerged from the both dissolution- and evaporation guided self-propulsion (see supplementary movies S1 and S2). The maximal thermal contrast observed in a course of the motion of grain and rotation of the rotor was $\Delta T \cong 0.2$ K. Thus, the soluto-capillary Marangoni flows are accompanied in the reported experimental situation with the thermo-capillary ones. Recall that the co-occurrence the soluto- and thermo-capillary flows takes place even for the famous phenomenon of "wine tears".[26] Let us estimate the contributions of thermo- and soluto-capillarity to the effect of self-propulsion. The total change in the surface tension $\Delta\gamma(T,c)$ is expressed as follows:

$$\Delta\gamma(T,c) = \Delta\gamma_1 + \Delta\gamma_2 = \left(\frac{\partial\gamma}{\partial c}\right)_T \Delta c + \left(\frac{\partial\gamma}{\partial T}\right)_c \Delta T \ , \tag{1}$$

where $\Delta\gamma_1 = \left(\frac{\partial\gamma}{\partial c}\right)_T \Delta c$ represents the contribution of soluto-capillarity, whereas $\Delta\gamma_2 = \left(\frac{\partial\gamma}{\partial T}\right)_c \Delta T$ is the term due to the thermo-capillarity. The value of the $\Delta\gamma_1$ was established in ref. 27 experimentally as $|\Delta\gamma_1| \cong 1.1 \times 10^{-3} \frac{J}{m^2}$. Assuming for the temperature gradient of the surface tension $\left|\left(\frac{\partial\gamma}{\partial T}\right)_c\right| \cong 17.7 \times 10^{-5} \frac{J}{K \times m^2}$ (see ref. 28); $\Delta T \cong 0.2 K$ we estimate $|\Delta\gamma_2| \cong 3.5 \times 10^{-5} \frac{J}{m^2}$. Thus, we conclude that the interrelation $|\Delta\gamma_1| \gg |\Delta\gamma_2|$ takes place and the motion of camphor boats is mainly due to the soluto-capillarity. However, the thermal effect may be of the primary importance for the understanding of the self-propulsion, when the strong exponential dependency of water viscosity is considered.[28] Thus, thermal waves play an essential role in breaking of the symmetry, resulting in the motion of the camphor boat.[23]

The thermal spirals evolved towards the walls of vessels during *ca* 0.5-1 s and afterwards remained stable (in other words stationary, see for example **Figure 2b,c**) during *c*a 10 s. The radial temporal displacement of the labeled water particle, located at the water/vapor surface, is shown in **Figure 3**. It is recognized from **Figure 3**, that at the initial stage of propagation, taking place ca 0.4 s, the radial displacement of the marker occurs with the constant velocity $v = \frac{dr}{dt} = const \cong 7.5 \frac{cm}{s}$. This means that at this stage the spiral is Archimedean one, described in the polar coordinates $(r, \theta)$ by the equation:

$$r(t) = \frac{v}{\omega}\theta(t) + c; \ \omega = const; c = const \tag{2}$$

It was demonstrated that the eventual Archimedean geometry of the spiral appearing under the Belousov-Zhabotinsky reactions may be independent of their chemical kinetic basis and arises from the symmetry considerations.[29] We are far from the exhaustive, quantitative explanation of the observed spiral, thermal waves. However, qualitative arguments will be useful for understanding of the phenomenon. What is the physico-chemical source of the observed thermal waves? One of these sources is enthalpy of the chemical reaction of dissolution of camphor by water. However, thermal waves were observed in the situation when the disc-like rotator, shown in **Figure 1b**, blocked the dissolution of camphor in water. Thus, it is plausible to relate the formation of the thermal waves to formation of the adsorbed layer of camphor molecules surrounding the self-propelled object. The spatially-temporal periodic nature of formation of this layer was discussed in ref. 30. The enthalpy of periodic adsorption, in turn, (consider that adsorption is always exothermic) may give rise to the reported thermal waves. The role of the adsorption of the volatile elements (which is camphor in our case) in the evolution of Marangoni flows was discussed in ref. 31. It is reasonable to relate the formation of the observed spiral thermal patterns to the capillary-gravity waves, generated by the self-propelled camphor boats.[32] Formation of Archimedian spiral patterns of capillary-gravity waves created by rotating floating bodies was reported in ref. 33. Similarity of waves created by floating bodies to the Cherenkov radiation was discussed in refs. 32-36. It should be emphasized that the formation of stable spiral thermal patterns was observed under the velocities of the self-propulsion $v \cong 2.0 \frac{cm}{s}$ which are much smaller than the threshold one $v_{min} = \sqrt[4]{\frac{4g\gamma}{\rho}} \cong 23 \frac{cm}{s}$ (where $\rho$ is the water density), necessary for formation of steady capillary-gravity waves under straight uniform self-propulsion.[33-36] This effect was explained in ref. 33, in which it was demonstrated that no velocity threshold exist for the steady spiral wave patterns, appearing under the circular motion of the self-propelled bodies, explored in our experiments.

Let us estimate dimensionless numbers, governing the evolving of spiral thermal waves, which are the Reynolds (Re) and Prandtl (Pr) numbers:

$$\text{Re} = \frac{\rho v L}{\eta}; \text{Pr} = \frac{c_p \eta}{\kappa}, \tag{3}$$

where $\eta, c_p$ and $\kappa$ are viscosity, specific heat and thermal conductivity of water at the conditions of the experiment respectively, $v$ and $L$ are the characteristic velocity and dimension correspondingly.[12,13] Assuming, $\rho = 10^3 \frac{kg}{m^3}; \eta \cong 9 \times 10^{-4} \text{Pa} \times \text{s}; c_p \cong 4.2 \times 10^3 \frac{J}{kg \times K}; \kappa \cong 0.6 \frac{W}{m \times K}; L \cong 10^{-3} - 10^{-2}$ m and $v \cong 10^{-2} - 10^{-1} \frac{m}{s}$ we estimate $Re \cong 10^1 - 10^2$ and $Pr \cong 6.3$. This means that the propagation of spiral waves occurs within the laminar regime; however, both momentum and heat dissipate through the fluid at about the same rate under evolving of the spiral thermal waves. The last consideration makes the exact quantitative analysis of the phenomenon extremely challenging.

To conclude we report the fascinating phenomenon of propagation of spiral thermal waves emerging from the motion of camphor-driven floating self-propelled objects. The soluto-capillarity mainly guides the self-propulsion. Thermo-capillarity in turn, gives rise to propagation of spiral Archimedean thermal waves, resembling those observed under Belousov-Zhabotinsky reaction.[6-9] We relate the observed thermal effect to the adsorption of camphor molecules at the water/vapor interface.[30,31] It seems reasonable to assume, that spiral thermal surface waves reflect the quasi-periodic distribution of camphor at the water/vapor interface.[30] The reported effect is important for understanding of the self-propulsion of the floating bodies, exploited recently for micro-robotics applications and ecologically friendly generation of electrical energy.[16-18, 37]

**Acknowledgements**

That authors are thankful to Mrs. Yelena Bormashenko for her kind help in preparing this manuscript. The authors are indebted to Professor Oleg Gendelman for extremely fruitful discussions.

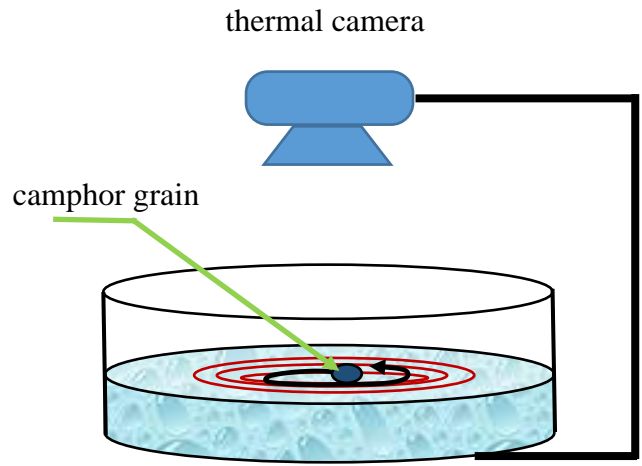

**Figure 1a**. Sketch of the experimental unit used for the study of spiral thermal waves arising from the self-propulsion of camphor grains. The self-propulsion is guided by the dissolution of camphor.

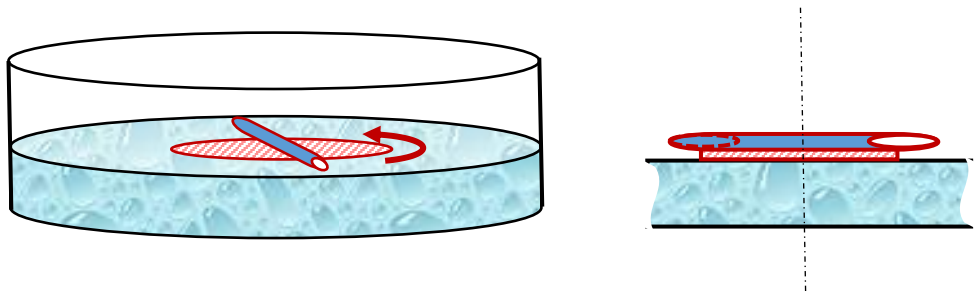

**Figure 1b**. Sketch of the special rotator enabling the study of the evaporation–guided self-propulsion is depicted. The polymer disk prevented direct contact of camphor with water.

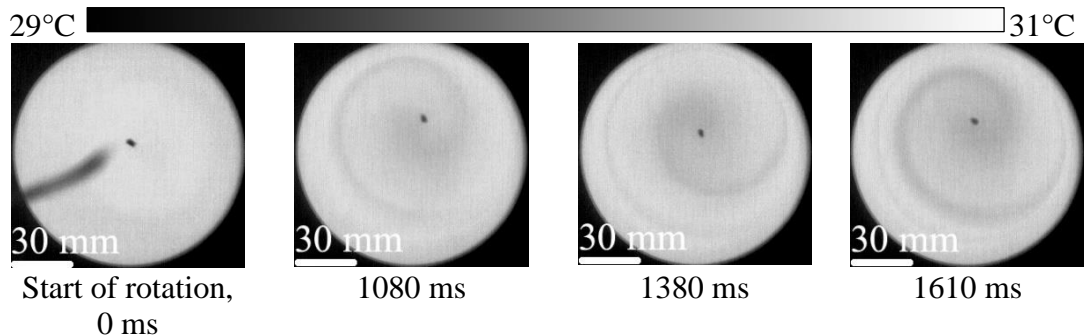

**Figure 2a**. The sequence of thermal images representing spiral thermal surface waves, generated under the dissolution-guided self-propulsion of camphor grains. Thermographic images were taken in the course of motion of the camphor grain. Brighter pixels correspond to the higher temperatures.

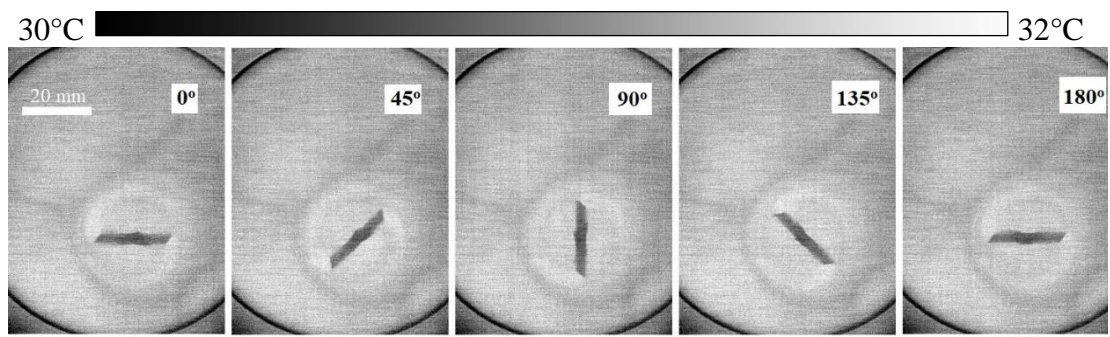

**Figure 2b**. The sequence of thermal images representing spiral thermal waves in the stationary regime, generated under the evaporation-guided self-propulsion of the rotor, shown in **Figure 1b.** Degrees illustrate the rotation of the polymer tubing filled with camphor. Thermographic images were taken in the course of the rotation. Brighter pixels correspond to the higher temperatures.

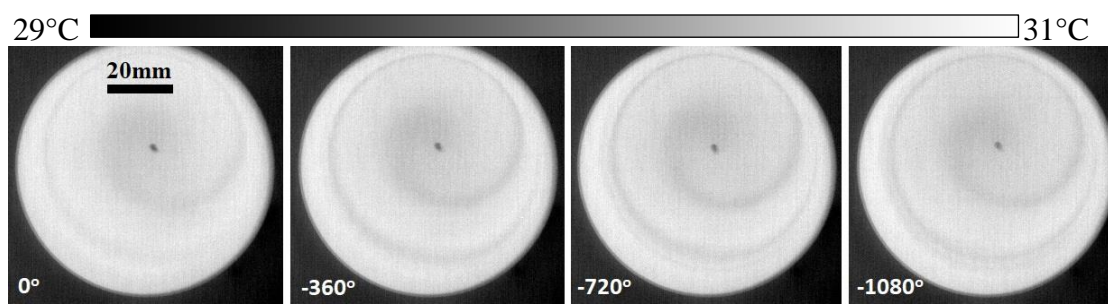

**Figure 2c**. Thermal images illustrating the stationary regime of rotation of the camphor sample is shown. Degrees illustrate the rotation of the camphor grain. The spiral thermal surface wave is clearly seen.

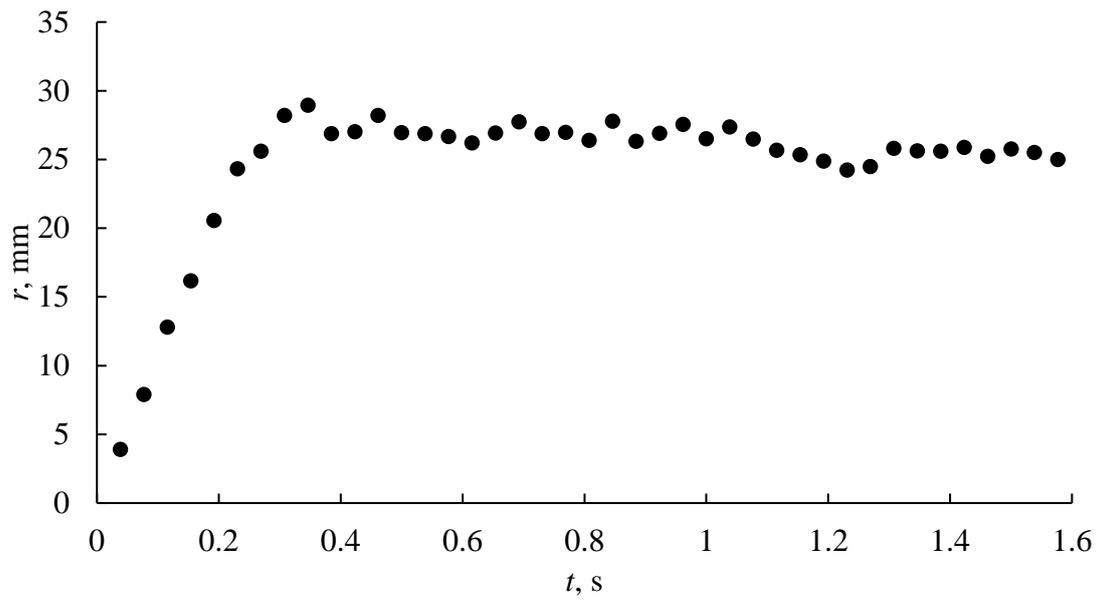

**Figure 3**. The temporal dependence of the radial displacement of the marker taken in the course of the spiral wave propagation is depicted.